\documentclass[superscriptaddress,a4paper]{revtex4}
\usepackage{amssymb}
\usepackage{geometry}
\usepackage{graphicx}

\begin{document}

\title{Hybridization Isotherms of DNA Microarrays 
and the Quantification of Mutation Studies}
\author{Avraham Halperin\thanks{Corresponding author: 
Fax: +33 4 38 78 56 91. Email: ahalperin@cea.fr}}
\author{Arnaud Buhot}
\affiliation{UMR 5819 (UJF, CNRS, CEA) DRFMC/SI3M, 
CEA Grenoble, 17 rue des Martyrs, 38054 Grenoble 
cedex 9, France}
\author{Ekaterina B. Zhulina}
\affiliation{Institute of Macromolecular Compounds 
of the Russian Academy of Sciences,
199004 St Petersburg, Bolshoy prospect 31, Russia}

\begin{abstract}
{\bf Background:} Diagnostic DNA arrays for detection of point mutations as
markers for cancer usually function in the presence of a large excess of
wild type DNA. This excess can give rise to false positives due to
competitive hybridization of the wild type target at the mutation spot. The
analysis of the DNA array data is typically qualitative aiming to establish
the presence or absence of a particular point mutation. Our theoretical
approach yields methods for quantifying the analysis so as to obtain the
ratio of concentrations of mutated and wild type DNA.

{\bf Method:} The theory is formulated in terms of the hybridization isotherms
relating the hybridization fraction at the spot to the composition of the
sample solutions at thermodynamic equilibrium. It focuses on samples
containing an excess of single stranded DNA and on DNA arrays with low
surface density of probes. The hybridization equilibrium constants can be
obtained by the nearest neighbor method.

{\bf Results:} Two approaches allow us to obtain quantitative results from the DNA
array data. In one the signal of the mutation spot is compared with that of
the wild type spot. The implementation requires knowledge of the saturation
intensity of the two spots. The second approach requires comparison of the
intensity of the mutation spot at two different temperatures. In this case
knowledge of the saturation signal is not always necessary.

{\bf Conclusions:} DNA arrays can be used to obtain quantitative results on the
concentration ratio of mutated DNA to wild type DNA in studies of somatic
point mutations.
\end{abstract}

\maketitle

\section{Introduction}

Cancers are attributed to accumulation of somatic mutations~\cite{Lodish}.
In turn, the mutated DNA can provide molecular markers of diagnostic utility~%
\cite{Kirk}. Among them are point mutations, involving a change in a single
base pair in the DNA, such as those occurring in the p53 and K-ras genes~%
\cite{Sidransky,Hibi,Marta}. The detection of such point mutations is useful
in screening for cancers~\cite{Prix} as well as typing the cancer in order
to optimize the treatment protocol~\cite{Golub}. DNA microarrays~\cite%
{Graves,Wang}, "DNA chips", are among the analytical techniques of proven
potential to this end~\cite{Prix,Livache}. The detection of somatic point
mutations is hampered by the presence of an excess of wild type DNA. This
favors hybridization of the wild type single stranded DNA (ssDNA) with
mismatched sequences resulting in false positives. A similar problem occurs
in analyzing single nucleotide polymorphism of pooled samples~\cite{Kirk}.
In the following we present a theoretical analysis of the errors introduced
by such mishybridization utilizing equilibrium thermodynamics. It suggests
methods of quantifying the detection of point mutations by DNA chips. In
particular, these methods allow to obtain the ratio of concentrations of
mutated and wild type DNA in the sample. This ratio is of diagnostic
interest and it provides a systematic method for the minimization of false
positives. The numerical calculations illustrating this approach are based
on recent DNA chip studies of point mutation in the K-ras gene~\cite%
{Prix,Livache}.

DNA chips consist of a support surface carrying "spots"~\cite{Graves,Wang}.
Each spot comprises numerous oligonucleotides of identical and known
sequence, "probes", that are terminally anchored to the surface. The spots
are placed in a checkered pattern such that each sequence is allocated a
unique site. Each probe hybridizes preferentially with a ssDNA containing a
complementary sequence referred to as "target". In a typical experiment the
DNA microarray is immersed in a solution containing a mixture of labelled
ssDNA fragments of unknown sequence. The presence of a given sequence is
signalled by the hybridization on the corresponding spot as monitored by
correlating the strength of the label signal with the position of the spot.
False positives can occur because each probe can also hybridize with a
mismatched sequence. When the DNA microarray is designed to investigate gene
expression it is possible to optimize the probe design in order to minimize
this effect~\cite{Lockhart,Li}. However, the implementation of this strategy
is more difficult for studies of point mutations. When studying somatic
point mutations the situation is further complicated by an excess of wild
type, non-mutated, DNA. Two observations substantiate this point. First,
solid tumors are heterogenous, containing a mixture both cancerous and
normal, stromal, cells. The cancer cells are a minority and the fraction of
mutated DNA obtained from homogenized tumor biopsy can be as small as $15\%$~%
\cite{Kirk}. The fraction is much smaller for noninvasive testing for early
stage cancers using body fluids such as urine~\cite{Marta}, serum~\cite{Hibi}
or stool~\cite{Sidransky,Prix}. Since the hybridization is controlled by the
mass action law, the excess of wild type sequence will typically contribute
to the hybridization on the spots allocated to the point mutations.
Accordingly, the ratio of intensities of different spots may not reflect the
ratio of concentrations of DNA species in the sample. This mishybridization
contribution will increase with the ratio of wild type DNA to mutated DNA
thus diminishing the efficacy of the early stage screening. A similar
situation is encountered in the analysis of pooled single nucleotide
polymorphism samples. This problem can be resolved by determining the
contribution of the wild type DNA to the signal of the mutation spots. As we
shall discuss, this is possible when three conditions are fulfilled: (i) the
hybridization is allowed to reach equilibrium. (ii) The equilibrium
hybridization isotherm, relating the hybridization fraction on the spot to
the sample composition, is of the Langmuir form i.e., $x/(1-x)=Kc$ where $x$
is the fraction of surface sites that bind a reactant of concentration $c$
and $K$ is the equilibrium constant of the reaction~\cite{Adamson}. This
regime is expected when the surface density of probes is sufficiently low~%
\cite{Halperin}. (iii) The sample contains an excess of ssDNA as is the case
when using asymmetric PCR amplification~\cite{Gyllensten,Gao} or following
digestion by lambda exonuclease~\cite{Higuchi}. Under these conditions the
fraction of correctly hybridized probes on a spot is obtainable yielding
also the concentrations of mutated and wild type DNA in the sample. Two
approaches are possible: (a) comparing the signals of two spots with probes
that match, respectively, the wild type and mutated sequences. (b) Comparing
the signals of the spot corresponding to the mutation of interest at two
different temperatures, $T_{1}$ and $T_{2}$. The first approach requires
measurement of the saturation signal of the two spots while for the second
this step may be eliminated. Importantly, the experimental set up reported
by Fotin et al~\cite{Fotin} satisfies the three conditions listed above and
allows to implement the two temperature approach.

Our analysis is based on the hybridization isotherms of DNA chips allowing
for the role of two types of competitive hybridization~\cite{Halperin}.
Competitive surface hybridization occurs when two different targets can
hybridize with the same probe. Competitive bulk hybridization takes place
when the target can hybridize with a complementary sequence in the bulk. The
second process is dominant when the samples are produced by symmetric PCR
amplification. When asymmetric PCR~\cite{Gyllensten,Gao} is used the sample
contains an excess of ssDNA which does not experience the effect of
competitive bulk hybridization. This situation can also be obtained by
digesting the product of symmetric PCR with lambda exonuclease~\cite{Higuchi}%
. Under these conditions the excess of ssDNA dominates the hybridization
with the probes and competitive surface hybridization becomes the major
source of error. In the general case the hybridization isotherm reflects the
electrostatic penalty incurred because each hybridization event increases
the charge of the probe layer~\cite{Halperin,Vainrub}. We will focus on
systems where these interactions are screened and the hybridization isotherm
assumes the Langmuir form~\cite{Adamson} which facilitates quantitative
analysis of the data. The use of PCR amplification renders the absolute
concentrations of DNA species meaningless. Our analysis is thus concerned
with the ratio $c_{m}/c_{w}$ of the excess concentrations of mutated ($c_{m}$%
) and wild form ($c_{w}$) of ssDNA.

\begin{figure}
\includegraphics[width=10cm]{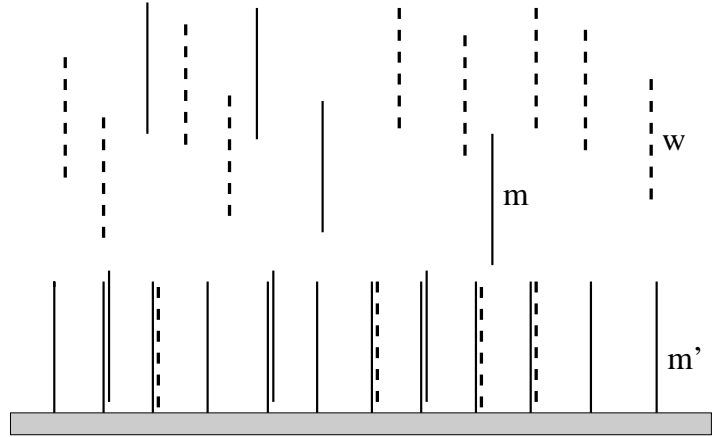}
\caption{{\bf Competitive Surface Hybridization.}
The $m^{\prime}$ probes at the mutation spot can hybridize with both the
complementary $m$ targets and the mismatched wild type $w$ targets. The
fraction of mishybridized, $m^{\prime}w$, probes is high when the $w$
targets are present in excess. Double stranded targets are not shown.}
\end{figure}

Our discussion focuses on the situation following amplification by
asymmetric PCR (Figure 1). It concerns two spots: a mutation spot carrying $%
m^{\prime }$ probes and a wild type spot carrying $w^{\prime }$ probes. The
sample solution comprises of a wild type ssDNA fragment $w$ and a mutated
fragment $m$ as well as their complementary fragments, $w^{\prime }$ and $%
m^{\prime }$. The number of $w$ and $m$ fragments is however larger. We
assume that all free $w^{\prime }$ and $m^{\prime }$ fragments hybridize
with the free complementary $w$ and $m$ ssDNA fragments. Our analysis
focuses on the excess of unhybridized $m$ and $w$ fragments whose
concentrations are denoted respectively by $c_{m}$ and $c_{w}$. We consider
the small spot limit, when the hybridization at the spot has a negligible
effect on the bulk concentrations. In this limit the double stranded DNA $%
ww^{\prime }$ and $mm^{\prime }$ pairs in the bulk do not affect the
hybridization equilibrium at the spot. In other words, the hybridization
isotherm relates $c_{m}$ and $c_{w}$ to the measured hybridization signal of
a spot. The targets are actually much larger than the probes. In the K-ras
studies we cite the target is comprised of 117~\cite{Livache} to 157~\cite%
{Prix} monomers (nucleotides) while the probes comprise of 13-14 monomers.
As we shall discuss, this "size asymmetry" mostly affects the boundary of
the Langmuir regime. Its effect on the hybridization isotherms of spots of
high probe density is beyond the scope of this article. In section II we
present the necessary background on the hybridization isotherm when
competitive surface hybridization is dominant. The following two section
discuss respectively the two-spot and two-temperature approaches for the
determination of $c_{m}/c_{w}$. Experimental considerations, the relation to
existing experimental studies as well as open questions are presented in the
Discussion.

\section{The Adsorption Isotherm}

The hybridization isotherm relates the equilibrium fraction of hybridized
probes on the spot to the composition of the sample~\cite{Halperin}. A
simple derivation of this isotherm is possible upon equating the rates of
hybridization and denaturation at the surface of the spot~\cite{Adamson}. In
the following we consider the isotherm obtained when competitive surface
hybridization is important. In particular, both the wild type sequence, $w$,
and the mutated one, $m$, can hybridize at the mutated $m^{\prime }$ spot.
Consider a spot carrying a total of $N_{T}$ $m^{\prime }$-probes in a sample
containing unhybridized $m$ and $w$ targets of concentrations $c_{m}$ and $%
c_{w}$ moles per liter respectively. The total hybridization reflects both
hybridization of the perfectly matched $m$ targets and of the mismatched $w$
ones. The fractions of $m^{\prime }m$ and $m^{\prime }w$ hybridized probes
are denoted respectively by $x$ and $y.$ The rate of denaturation of
mismatched $m^{\prime }w$ probes is $k_{dm}yN_{T}$ while the rate of
hybridization of $m^{\prime }$ probes with $w$ targets is $%
k_{hm}c_{w}(1-x-y)N_{T}$ where $k_{dm}$ and $k_{hm}$ are the corresponding
rate constants. When electrostatic interactions within the layer are
strongly screened, the rate constants do not depend on the probe density and
the fraction of hybridized probes. At equilibrium the two rates are equal, $%
k_{dm}yN_{T}=k_{hm}c_{w}(1-x-y)N_{T}$ leading to%
\begin{equation}
\frac{y}{\left( 1-x-y\right) }=c_{w}\frac{k_{hm}}{k_{dm}}=c_{w}K_{m^{\prime
}w}=c_{w}\exp (-\Delta G_{m^{\prime }w}^{0}/RT).  \label{II1}
\end{equation}%
where $k_{hm}/k_{dm}=K_{m^{\prime }w}$ is the equilibrium constant~\cite%
{Moore}. Similarly, the two rates for the perfectly matched $m^{\prime }m$
probes are $k_{dp}xN_{T}$ and $k_{hp}c_{m}(1-x-y)N_{T}$. Their equality at
equilibrium, $k_{dp}xN_{T}=$ $k_{hp}c_{m}(1-x-y)N_{T}$ leads to 
\begin{equation}
\frac{x}{\left( 1-x-y\right) }=c_{m}\frac{k_{hp}}{k_{dp}}=c_{m}K_{m^{\prime
}m}=c_{m}\exp (-\Delta G_{m^{\prime }m}^{0}/RT).  \label{II2}
\end{equation}%
Here $K_{m^{\prime }w}$ and $K_{m^{\prime }m}$ are respectively the
equilibrium constants of hybridization between $m^{\prime }$ probes and $w$
and $m$ targets, $\Delta G_{m^{\prime }w}^{0}$ and $\Delta G_{m^{\prime
}m}^{0}$ are the corresponding standard Gibbs free energies, $T$ is the
absolute temperature and $R$ is the gas constant. This kinetic derivation
recovers the results of a rigorous thermodynamic analysis~\cite{Halperin}
with one caveat: The molar concentrations should be replaced by
dimensionless activities $a_{i}=\gamma c_{i}$ where $\gamma $ is the
activity coefficient~\cite{Moore}. Since $\gamma \rightarrow 1$ as $%
c_{i}\rightarrow 0$ we will retain expressions used above noting that the $%
c_{i}$ are dimensionless having the numerical value of the molar
concentrations of the $i^{th}$ species. The two isotherms, (\ref{II1}) and (%
\ref{II2}), are not helpful because PCR amplification of a given sample does
not allow for different labels of the $m$ and $w$ targets. Since both are
labeled identically, the measurement yields the total hybridization fraction 
$\theta =x+y$ rather than $x$ and $y$ separately. Accordingly, the observed
isotherm for the $m^{\prime }$ spot is 
\begin{equation}
\Omega _{m^{\prime }}=\frac{\theta _{m^{\prime }}}{1-\theta _{m^{\prime }}}%
=K_{m^{\prime }m}c_{m}+K_{m^{\prime }w}c_{w}  \label{II3}
\end{equation}%
as obtained by summing (\ref{II1}) and (\ref{II2}). The fraction of
mishybridized probes on the $m^{\prime }$ spot is 
\begin{equation}
P_{m^{\prime }}=\frac{y_{m^{\prime }}}{\theta _{m^{\prime }}}=\frac{%
K_{m^{\prime }w}c_{w}}{K_{m^{\prime }m}c_{m}+K_{m^{\prime }w}c_{w}}.
\label{II4}
\end{equation}%
Here, and in the following, the subscript $i=m^{\prime },w^{\prime }$ of $%
\Omega _{i},$ $P_{i}$ etc. identifies the spot. $P_{m^{\prime }}=1/2$ when $%
c_{w}=c_{m}K_{m^{\prime }m}/K_{m^{\prime }w}$ and $P_{m^{\prime }}>1/2$ for $%
c_{w}>c_{m}K_{m^{\prime }m}/K_{m^{\prime }w}$. In the case of interest, when 
$c_{w}\gg c_{m}$, $P_{m^{\prime }}$ of the $m^{\prime }$ spot is large while 
$P_{w^{\prime }}$ of the $w^{\prime }$ spot 
\begin{equation}
P_{w^{\prime }}=\frac{y_{w^{\prime }}}{\theta _{w^{\prime }}}=\frac{%
K_{w^{\prime }m}c_{m}}{K_{w^{\prime }w}c_{w}+K_{w^{\prime }m}c_{m}}
\label{II5}
\end{equation}%
is always small. The values of $P_{m^{\prime }}$ (Figure 2) for a typical
situation (Table 1) confirm that competitive surface hybridization is
important only when the competing species is present in large excess. As
noted earlier, this is the case in studies of somatic point mutations when
the wild type ssDNA is a majority component. The wild type ssDNA then
contributes to the hybridization on all the mutation spots. In contrast, the
concentrations of the different mutated ssDNA are much smaller. As a result
their contribution to the hybridization on the $m^{\prime }$ spot, and on
other mutation spots, is negligible. This observation justifies limiting the
analysis to the competition between two species, $m$ and $w.$

\begin{table}
\begin{tabular}{ccccc}
$i^{\prime }j$ & $\Delta H_{i^{\prime }j}^{0}kcal/mole$ & $\Delta
S_{i^{\prime }j}^{0}cal/mole\cdot \deg $ & $\Delta G_{i^{\prime
}j}^{0}(37^{\circ }C)kcal/mole$ & $K_{i^{\prime }j(37^{\circ }C)}$ \\ 
$m^{\prime }m$ & $-99.40$ & $-264.29$ & $-17.43$ & $2.11\times 10^{12}$ \\ 
$m^{\prime }w$ & $-78.20$ & $-214.12$ & $-11.79$ & $2.19\times 10^{8}$ \\ 
$w^{\prime }w$ & $-100.80$ & $-270.55$ & $-16.89$ & $8.75\times 10^{11}$ \\ 
$w^{\prime }m$ & $-93.50$ & $-253.59$ & $-14.85$ & $3.17\times 10^{10}$%
\end{tabular}
\caption{The thermodynamic parameters utilized in the numerical calculations
correspond to the Alanine 12 and wild type probes utilized by Prix et al~%
\cite{Prix}: $m^{\prime} = AGCTG\overline{C}TGGCGTA$, $m=TCGACGACCGCAT$, $%
w^{\prime}= CTGGTGGCGTAGG$, $w=GACCACCGCATCC$ as calculated from the nearest
neighbor model~\cite{HYTHERTM} for $1MNaCl$. Since the targets are longer
than the probes two dangling ends are invoked.}
\end{table}

\begin{figure}
\includegraphics[width=10cm]{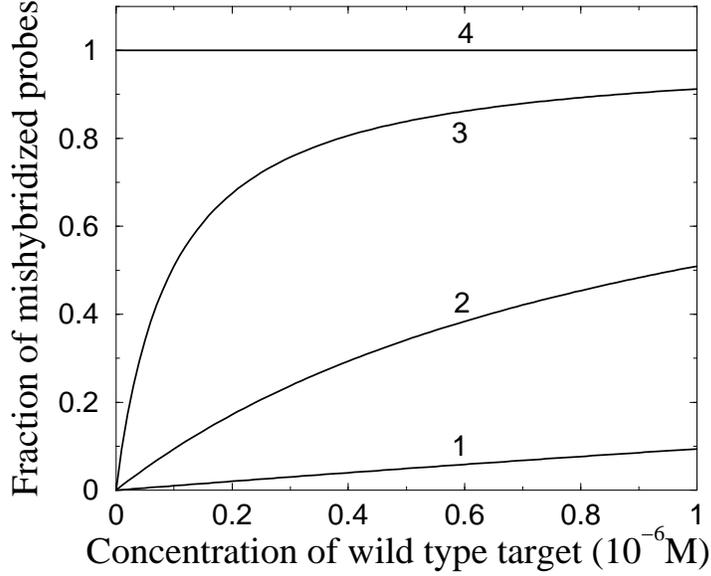}
\caption{{\bf The fraction of mishybridized probes on 
the mutation spot as function of the concentration of 
the wild type target.} $P_{m^{\prime}}$ vs. $c_{w}$ 
curves calculated at $T = 47^{\circ}C$ for (1) 
$c_{m}= 10^{-9}M$, (2) $c_{m}=10^{-10}M$, (3) $c_{m}=
10^{-11}M$ and (4) $c_m = 0$.}
\end{figure}

When competitive hybridization is negligible, equation (\ref{II3}) reduces
to $\Omega _{m^{\prime }}=K_{m^{\prime }m}c_{m}$ and the corresponding
isotherm for the wild type is $\Omega _{w^{\prime }}=K_{w^{\prime }w}c_{w}$.
In such cases $c_{m}/c_{w}$ can be determined from%
\begin{equation}
\frac{\Omega _{m^{\prime }}}{\Omega _{w^{\prime }}}= \frac{K_{m^{\prime }m}}{%
K_{w^{\prime }w}}\frac{c_{m}}{c_{w}}  \label{II5a}
\end{equation}
This situation can be realized in gene expression studies with proper probe
design and in the absence of overwhelming excess of one target species. Note
that $\theta $ of the mutation spot can be tuned to be low, $\theta \ll 1$
by adjusting the hybridization temperature. In such regimes one may invoke a
"weak spot" approximation 
\begin{equation}
\Omega_{i}\simeq \theta_{i}  \label{II6}
\end{equation}
As we shall discuss, when applicable this simplifies the analysis of the two
temperature experiments. Finally, in most cases the signal of the spot $i$
at equilibrium, $I_{i}$ is proportional to $\theta_{i}N_{T}$ and 
\begin{equation}
\theta _{i}=\frac{I_{i}}{I_{i}^{\max }}  \label{II7}
\end{equation}
where $I_{i}^{\max }$ is the saturated signal of the $i$ spot as obtained
upon equilibration with a concentrated solution of the perfectly matched
target. It is important to note that $I_{i}^{\max}$ can depend on
temperature.

The melting temperatures of double stranded DNA are used as design criterion
for probes~\cite{Li,Prix}. It is thus useful to note that competitive
surface hybridization affects the effective melting temperature, $T_{M}$. $%
T_{M}$ is defined by the condition $\theta _{i}=1/2$ or $\Omega _{i}=1$. In
the general case $T_{M}$ depends on both $c_{m}$ and $c_{w}$ i.e., $%
T_{M}=T_{M}(c_{m},c_{w})$. Note that there are two targets that can
hybridize with the same probe. This situation, involving three ssDNA species
and two different dsDNA, is different from the one invoked in the definition
of the melting temperature of a dsDNA. In this last case two complementary
ssDNA hybridize to form one type of dsDNA and the total number of species is
three rather than five. Furthermore, we are considering the small spot limit
where the hybridization with the probes has negligible effect on the bulk
composition. In the absence of competitive surface hybridization $T_{M}$
reduces to its familiar forms: for $c_{w}=0,$ this leads $T_{m^{\prime
}m}(c_{m})=\Delta H_{m^{\prime }m}^{0}/(\Delta S_{m^{\prime }m}^{0}+R\ln
c_{m})$ while for $c_{m}=0$ it leads to $T_{m^{\prime }w}(c_{w})=\Delta
H_{m^{\prime }w}^{0}/(\Delta S_{m^{\prime }w}^{0}+R\ln c_{w})$. It is not
possible to obtain an explicit analytical expression for $T_{M}(c_{m},c_{w})$%
. However, in the regime of interest, $c_{m}/c_{w}\ll 1$ the melting
temperature is well approximated by the first two terms of the Taylor
expansion in $c_{m}$ around $c_{m}=0$ i.e., $T_{M}(c_{m},c_{w})\simeq
T_{m^{\prime }w}+c_{m}\frac{\partial T_{M}}{\partial c_{m}}|_{c_{m}=0}$ thus
specifying the melting temperature at the mutation spot 
\begin{equation}
T_{M}(c_{m},c_{w})=T_{m^{\prime }w}(c_{w})+c_{m}\frac{K_{m^{\prime
}m}(T_{m^{\prime }w})}{|\Delta H_{m^{\prime }w}^{0}|}RT_{m^{\prime }w}^{2}
\label{II7b}
\end{equation}%
Importantly, in this regime $T_{M}$ increases steeply with $c_{m}$ (Figure
3) and its initial value, for $c_{m}=0$ is $T_{m^{\prime }w}$ rather than $%
T_{m^{\prime }m}$. In marked distinction, for $c_{m}/c_{w}\gtrsim 1$ the
effect of the competitive surface hybridization is negligible and $%
T_{M}\approx T_{m^{\prime }m}(c_{m})$ with a weak logarithmic dependence on $%
c_{m}$.

\begin{figure}
\includegraphics[width=10cm]{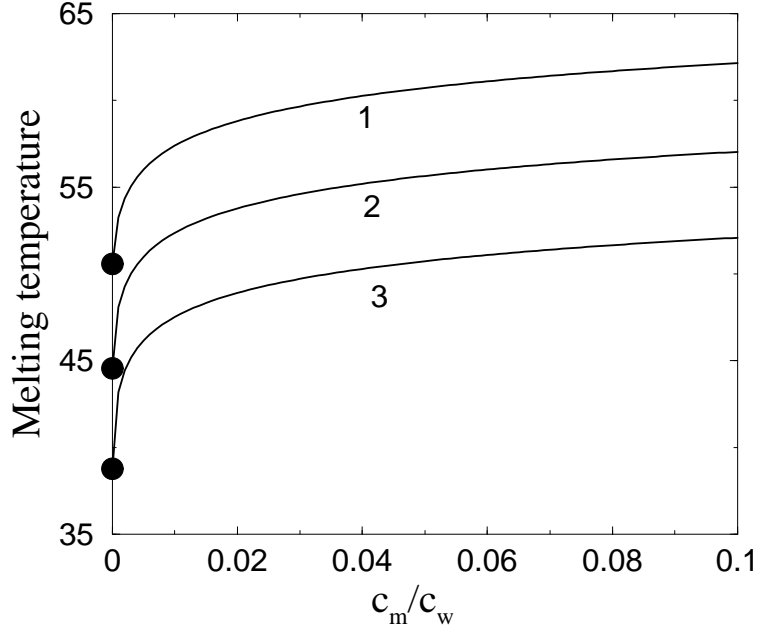}
\caption{{\bf The dependence of the melting temperature 
of the mutation spot on the concentration of the mutation 
target.} $T_{M}(c_w,c_m)$ in $^{\circ}C$ is plotted vs. 
$c_{m}/c_{w}$ for various $c_{m}+c_{w}=c$: (1) $c = 10^{-6}M$ 
(2) $c = 10^{-7}M$ and (3) $c = 10^{-8}M$. Dots correspond 
to the melting temperatures $T_{m^{\prime}w}(c_w)$.}
\end{figure}

When the probe density is relatively high and the salt concentration is
sufficiently low, it is necessary to allow for the electrostatic penalty
incurred upon hybridization because of the extra charge deposited at the
probe layer. In this regime the hybridization isotherms assume a different
form exemplified by 
\begin{equation}
\Omega _{m^{\prime }}=\frac{\theta _{m^{\prime }}}{1-\theta _{m^{\prime }}}%
=\left( K_{m^{\prime }m}c_{m}+K_{m^{\prime }w}c_{w}\right) \exp \left[
-\Gamma _{m^{\prime }}(1+\theta _{m^{\prime }})\right] .  \label{II8}
\end{equation}%
For the case of probes and targets of equal and short length $\Gamma
_{m^{\prime }}=$\textit{constant}$\times \sigma _{m^{\prime }}$ where $%
\sigma _{m^{\prime }}$ is the charge density of the unhybridized $m^{\prime
} $ spot and the constant is set by the ionic strength and the length of the
probe~\cite{Halperin}. In the following we consider the opposite limit, when
the electrostatics interactions are screened and the chains do not interact.
The hybridization is then described by a simple Langmuir isotherm that is,
the case of $\Gamma _{i}\simeq 0$. Importantly, in studies of point
mutations the targets are typically longer than the probes. Usually, the
hybridization site is situated roughly in the middle of the target.
Accordingly, a hybridized probe incorporates two unhybridized target
sections of similar length. The span of such segment is $R_{0}\approx
(nal)^{1/2}$ where $n$ is the number of bases in the section, $a$ is the
base size and $l$ is the persistence length. Typical values are $a\approx
0.6nm$ and $l\approx 0.75nm$~\cite{Smith}. The Langmuir regime occurs when
the area per probe, $\Sigma $, exceeds $R_{0}^{2}$ thus ensuring that the
hybridized probes do not interact. For the target-probe pairs considered
this condition is satisfied when $\Sigma \gtrsim 65nm^{2}$ or less than $%
1.5\times 10^{4}$ grafted probes per $\mu m^{2}$. Operation in this range
allows us to benefit from the absence of the non-linear behavior introduced
by the $\exp \left[ -\Gamma _{m^{\prime }}(1+\theta _{m^{\prime }})\right] $
term.

The hybridization isotherm is only applicable at equilibrium. In turn, this
implies two conditions: stationary signal and path independence i.e.,
independence of the preparation method and sample history. The reported
times required to attain stationary signal vary between minutes~\cite{Fotin}
to 14 hours~\cite{Peterson,Peterson2002}. Importantly, the equilibration
time may depend on the probe density, hybridization conditions, the
equilibrium $\theta _{i}$ etc. Path independence requires that the measured
signal at equilibrium will not change with the thermal history (heating and
cooling cycles). As stated earlier, these conditions are satisfied by the
system of Fotin et al~\cite{Fotin}. Under conditions of thermodynamic
equilibrium the quantitative methods of analysis we describe below are
applicable. However, these methods require knowledge of $\theta_{i}$ and $%
\Omega_{i}$. In turn, to obtain $\theta_{i}$ and $\Omega_{i}$ it is
necessary to ascertain the saturation value of the signal of the $i$ spot, $%
I_{i}^{\max}$ or, equivalently, the $N_{T}$ of the $i$ spot.

Numerical implementation of the analysis we describe requires knowledge of
the equilibrium constants $K_{m^{\prime }m}$ and $K_{m^{\prime }w}$ as well
as $K_{w^{\prime }m}$ and $K_{w^{\prime }w}$. These can be best obtained
experimentally from the hybridization isotherm of a spot in contact with
single component samples. However, this approach is time consuming when a
number of equilibrium constants is required. Fortunately, in the Langmuir
regime it is reasonable to approximate the equilibrium constants by their
bulk values. The nearest neighbor model, with the unified set of parameters
compiled by Santa Lucia and collaborators~\cite%
{SantaLucia98,Peyret,Bommarito}, allows to calculate $\Delta H^{0}$ and $%
\Delta S^{0}$ for the hybridization of perfectly matched oligonucleotide
pairs as well as for pairs containing a single mismatch. In the numerical
calculations we use probe target pairs incorporating the 12$^{th}$ and 13$%
^{th}$ K-ras codons. The $\Delta H^{0}$ and $\Delta S^{0}$ values are
obtained by use of HYTHERTM software implementing the nearest neighbor
approach~\cite{HYTHERTM}. The identity of the probes and targets considered
in the numerical calculations as well as the corresponding standard
enthalpies, entropies are listed in Table 1. The table also lists the free
energies of formation and the equilibrium constants of hybridization at $%
37^{0}C$. The hybridization temperatures considered are chosen in view of
the conditions in the cited experiments. Fotin et al~\cite{Fotin} studied
the hybridization at the range of $T=-20^{0}C$ to $T=60^{0}C$. In the K-ras
studies the hybridization temperatures were $37^{0}C$~\cite{Prix} and $%
50^{0}C$~\cite{Livache}. The $c_{m}$ values in the numerical examples vary
around $3\cdot 10^{-10}M$, while the concentration of the control target $%
c_{w}$ is $10^{2}$ to $10^{3}$ larger~\cite{Prix}.

\section{The Two Spot Approach}

When two spots carrying respectively wild type, $w^{\prime },$ and mutation, 
$m^{\prime },$ probes are placed in contact with a solution of $w$ and $m$
targets, the corresponding isotherms are%
\begin{equation}
\Omega _{w^{\prime }}=c_{w}K_{w^{\prime }w}+c_{m}K_{w^{\prime }m}
\label{III1}
\end{equation}
\begin{equation}
\Omega _{m^{\prime }}=c_{w}K_{m^{\prime }w}+c_{m}K_{m^{\prime }m}
\label{III2}
\end{equation}%
where $\Omega _{i}=\theta _{i}/(1-\theta _{i})$ of spot $i=w^{\prime
},m^{\prime }$ and $\theta _{i}$ is the corresponding fraction of hybridized
probes, irrespective of their identity (mismatched or perfectly matched).
These two equations immediately determine $c_{w}$ and $c_{m}$%
\begin{equation}
c_{w}=\frac{\Omega _{w^{\prime }}K_{m^{\prime }m}-\Omega _{m^{\prime
}}K_{w^{\prime }m}}{K_{w^{\prime }w}K_{m^{\prime }m}-K_{w^{\prime
}m}K_{m^{\prime }w}}  \label{III3}
\end{equation}%
\begin{equation}
c_{m}=\frac{\Omega _{m^{\prime }}K_{w^{\prime }w}-\Omega _{w^{\prime
}}K_{m^{\prime }w}}{K_{w^{\prime }w}K_{m^{\prime }m}-K_{w^{\prime
}m}K_{m^{\prime }w}}.  \label{III4}
\end{equation}
Accordingly 
\begin{equation}
\frac{c_{m}}{c_{w}}=\frac{\Omega _{m^{\prime }}K_{w^{\prime }w}-\Omega
_{w^{\prime }}K_{m^{\prime }w}}{\Omega _{w^{\prime }}K_{m^{\prime }m}-\Omega
_{m^{\prime }}K_{w^{\prime }m}}=\frac{K_{w^{\prime }w}-\alpha K_{m^{\prime
}w}}{\alpha K_{m^{\prime }m}-K_{w^{\prime }m}}  \label{III5}
\end{equation}
where, in the regime considered, $\alpha =\Omega _{w^{\prime }}/\Omega
_{m^{\prime }}\gg 1$. The range of $\alpha $ values varies between $\alpha
_{\max }=K_{w^{\prime }w}/K_{m^{\prime }w}\gg 1$, corresponding to $c_{m}=0$
and $\alpha _{\min }=K_{w^{\prime }m}/K_{m^{\prime }m}\ll 1$ when $c_{w}=0$.
In the realistic limit of $\alpha K_{m^{\prime }w}\ll K_{w^{\prime }w}$ this
expression reduces to $c_{m}/c_{w}\approx K_{w^{\prime }w}/\alpha
K_{m^{\prime }m}$ that is%
\begin{equation}
\frac{c_{m}}{c_{w}}=\frac{K_{w^{\prime }w}}{K_{m^{\prime }m}}\frac{%
I_{m^{\prime }}}{I_{w^{\prime }}}\frac{I_{w^{\prime }}^{\max }-I_{w^{\prime
}}}{I_{m^{\prime }}^{\max }-I_{m^{\prime }}}  \label{III6}
\end{equation}%
where $I_{m^{\prime }}$ and $I_{w^{\prime }}$ are the measured intensities
of the $m^{\prime }$ and $w^{\prime }$ spots while $I_{m^{\prime }}^{\max }$
and $I_{w^{\prime }}^{\max }$ are the corresponding saturation values. Thus,
the implementation of this approach requires knowledge of $N_{T}$, or
equivalently $I_{\max },$ for the two spots. \ In the general case it is
necessary to know four equilibrium constants while in the simplest case, of $%
\alpha K_{m^{\prime }w}\ll K_{w^{\prime }w}$, knowledge of $K_{m^{\prime }m}$
and $K_{w^{\prime }w}$ is sufficient.

\begin{figure}
\includegraphics[width=10cm]{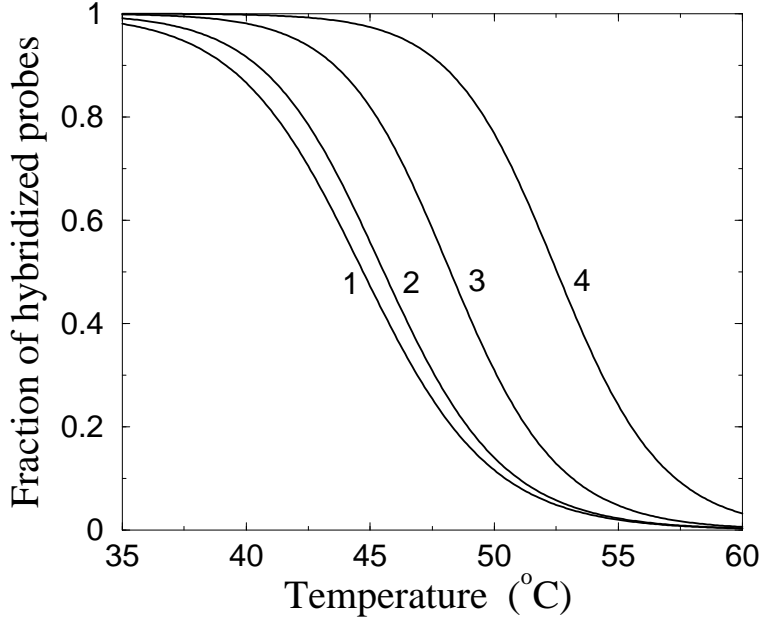}
\caption{{\bf Melting curves of the mutation spot.} 
$\theta_{m^{\prime}}$ vs. temperature curves are plotted 
for $c_{m}+c_{w} = c = 10^{-7}M$ and (1) $c_{m}/c_{w}=0$, 
(2) $c_{m}/c_{w}=10^{-4}$, (3) $c_{m}/c_{w}=10^{-3}$ and
(4) $c_{m}/c_{w}= 10^{-2}$. For a given $c_w \simeq c$ the 
melting temperature and the saturation temperature increase 
with $c_m/c_w$.}
\end{figure}

\begin{figure}
\includegraphics[width=10cm]{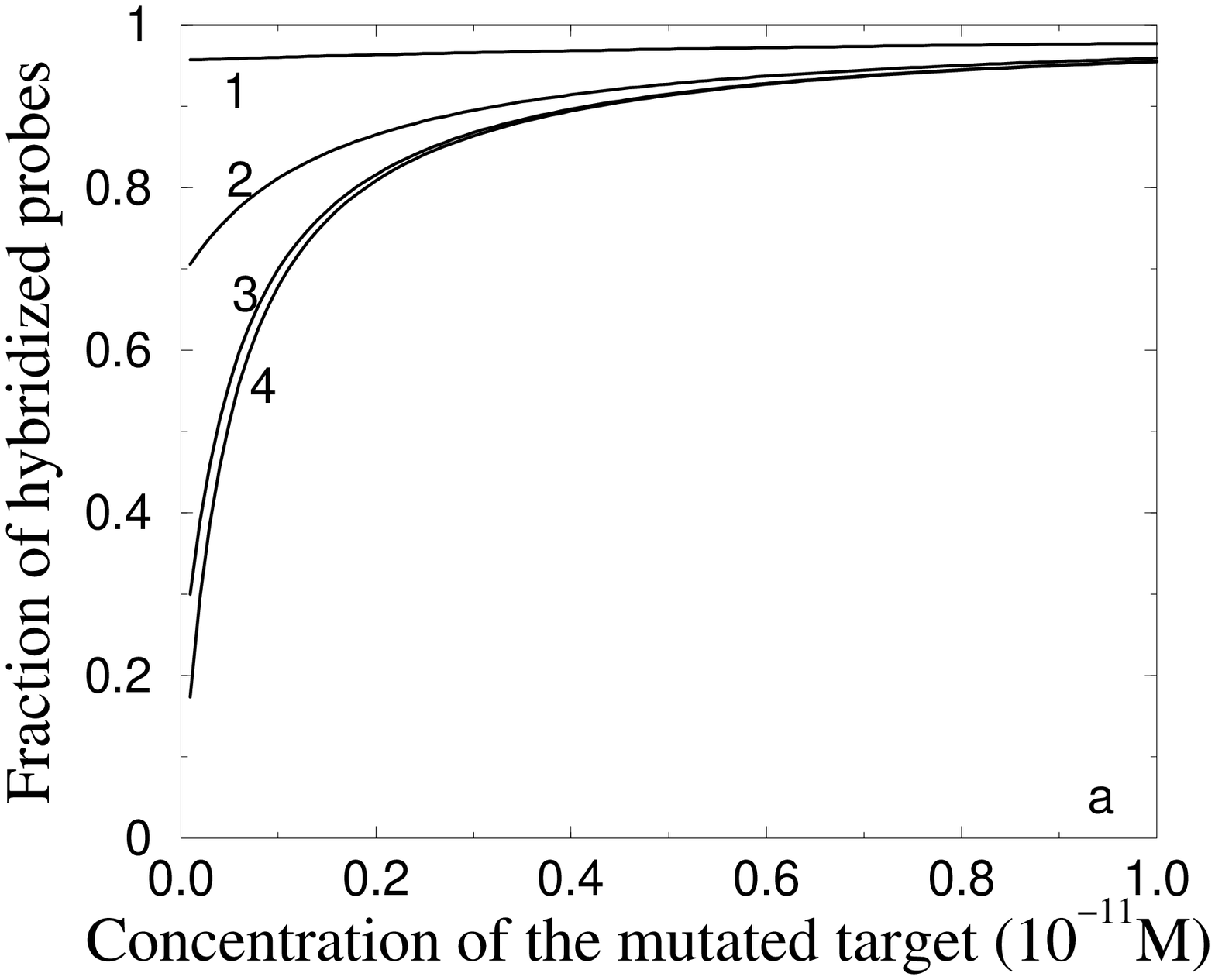}
\includegraphics[width=10cm]{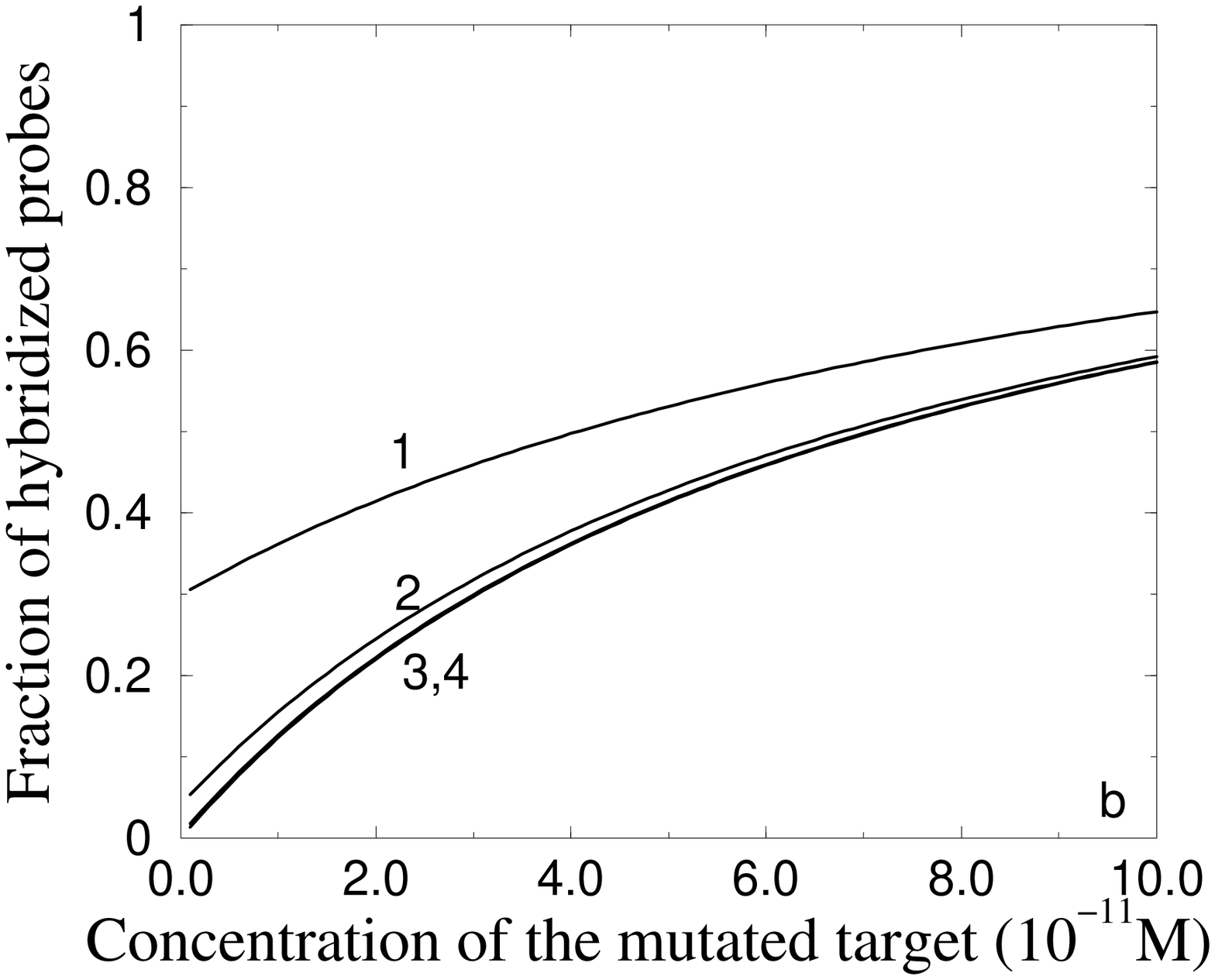}
\caption{{\bf The hybridization isotherm of the mutation spot.}
$\theta _{m^{\prime}}$ vs. $c_{m}$ curves are plotted at 
(a) $T=37^{\circ}C$ (b)$T=47^{\circ}C$ for (1) $c_{w} 
= 10^{-7}M$ (2) $c_{w} = 10^{-8}M$ (3) $c_{w} = 10^{-9}M$
and (4) $c_w = 0$. The saturation regime can be avoided by 
increasing the hybridization temperature.}
\end{figure}

The two spots approach is feasible when the values of $\theta_{m^{\prime}}$
and $\theta_{w^{\prime}}$ for different sets of $c_{m}$ and $c_{w}$ are
distinguishable. In practical terms this imposes two requirements. First, it
is necessary to avoid the saturation regimes of the melting curve (Figure 4)
and the hybridization isotherm (Figure 5). Second, the $\theta_{i^{\prime}} $
values corresponding to the sample composition must be large enough in
comparison with the experimental errors. One may optimize the performance by
tuning the hybridization temperature $T$: Increasing $T$ lowers the
hybridization degree thus preventing saturation at the price of weaker $%
\theta_{i^{\prime}}$ and a higher noise to signal ratio.

\begin{figure}
\includegraphics[width=10cm]{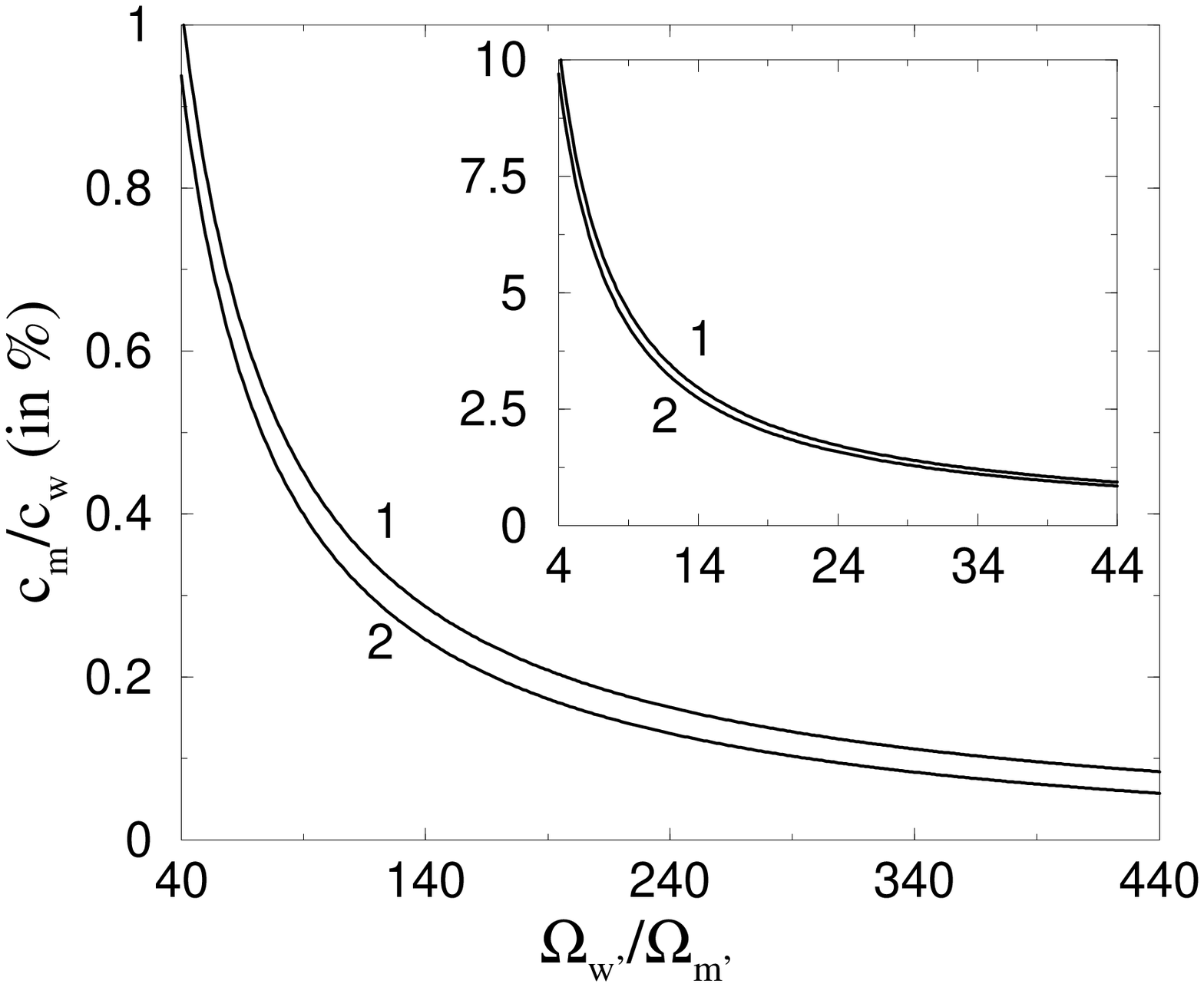}
\caption{{\bf The ratios of concentrations obtainable 
from a "2-spot" experiment.} $c_{m}/c_{w}$ vs. 
$\Omega_{w^{\prime}}/\Omega_{m^{\prime}}$ curves
are plotted at (1) $T = 37^{\circ}C$ and (2)
$T= 47^{\circ}C$ in the range $0.05 \% \leq c_m/c_w 
\leq 1 \%$. Inset: Range $1 \% \leq c_m/c_w \leq 10 \%$.}
\end{figure}

The operational conditions are thus determined by two inputs: The typical $%
c_w$ of the amplicon and the desired range of $c_m/c_w$. Once these two
parameters are specified it is possible to calculate the relevant melting
curves and hybridization isotherms in order to choose the hybridization
temperature. As noted before (Eq.~\ref{III5}), $c_m/c_w$ is extracted from $%
\alpha = \Omega_{w^{\prime}}/\Omega_{m^{\prime}}$. The temperature
dependence of $\alpha$ is relatively weak (Figure 6). The accessible range
of $\alpha$ depends however on the detection limit of $\theta_{m^{\prime}}$
which sets a lower bound of $\theta_{m^{\prime}}$, $\theta_{m^{\prime}}(min)$%
. Since $\Omega_{w^{\prime}} \leq 1$ the maximal range of $\alpha$ is
roughly $1/\theta_{m^{\prime}}(min)$.

\section{The Two Temperature Approach}

An alternative approach involves equilibration of the DNA chip with the
biological sample at two different temperatures, $T_{1}$ and $T_{2}$.
Focusing on the $m^{\prime}$ spot, we have 
\begin{equation}
\Omega_{m^{\prime}}(T_{1})=c_{w}K_{m^{\prime}w}(T_{1})
+c_{m}K_{m^{\prime}m}(T_{1}),  \label{IV1}
\end{equation}
\begin{equation}
\Omega_{m^{\prime}}(T_{2})=c_{w}K_{m^{\prime}w}(T_{2})
+c_{m}K_{m^{\prime}m}(T_{2}).  \label{IV2}
\end{equation}
Solving equations (\ref{IV1}) and (\ref{IV2}) leads to 
\begin{equation}
c_{w}=\frac{\Omega_{m^{\prime}}(T_{2})K_{m^{\prime}m}(T_{1})-
\Omega_{m^{\prime}}(T_{1})K_{m^{\prime}m}(T_{2})}{K_{m^{\prime}w}
(T_{2})K_{m^{\prime}m}(T_{1})-K_{m^{\prime}w}(T_{1}) K_{m^{\prime}m}(T_{2})},
\label{IV3}
\end{equation}
\begin{equation}
c_{m}=\frac{\Omega_{m^{\prime}}(T_{1})K_{m^{\prime}w}(T_{2})
-\Omega_{m^{\prime}}(T_{2})K_{m^{\prime}w}(T_{1})}{K_{m^{\prime}w}
(T_{2})K_{m^{\prime}m}(T_{1})-K_{m^{\prime}w}(T_{1})K_{m^{\prime}m} (T_{2})}
\label{IV4}
\end{equation}
and 
\begin{equation}
\frac{c_{m}}{c_{w}}=\frac{\Omega _{m^{\prime }}(T_{1})K_{m^{\prime
}w}(T_{2})-\Omega _{m^{\prime }}(T_{2})K_{m^{\prime }w}(T_{1})}{\Omega
_{m^{\prime }}(T_{2})K_{m^{\prime }m}(T_{1})-\Omega _{m^{\prime
}}(T_{1})K_{m^{\prime }m}(T_{2})}.  \label{IV5}
\end{equation}%
To simplify this equation it is convenient to express $\frac{\Delta
H_{m^{\prime }w}}{RT_{2}}$ as $\frac{\Delta H_{m^{\prime }w}}{RT_{1}}(1- 
\frac{\Delta T}{T_{1}+\Delta T})$ where $T_{2}\ =T_{1}+\Delta T$. Upon
defining $\lambda =\Omega _{m^{\prime }}(T_{2})/\Omega _{m^{\prime }}(T_{1})$
we obtain $c_{m}/c_{w}$ in the form 
\begin{equation}
\frac{c_{m}}{c_{w}}=\frac{K_{m^{\prime }m}(T_{1})}{K_{m^{\prime }w}(T_{1})}%
\frac{\exp \left( \frac{\Delta H_{m^{\prime }w}}{RT_{1}}\frac{\Delta T}{%
T_{1}+\Delta T}\right) -\lambda }{\lambda -\exp \left( \frac{\Delta
H_{m^{\prime }m}}{RT_{1}}\frac{\Delta T}{T_{1}+\Delta T}\right) }.
\label{IV6}
\end{equation}%
Equation (\ref{IV6}) can be simplified further when $\Delta T/T_{1}\ll 1$
thus allowing to approximate $\frac{\Delta T}{T_{1}+\Delta T}\approx \frac{
\Delta T}{T_{1}}$. In the weak spot limit $\lambda =\frac{I_{m^{\prime
}}(T_{2})}{I_{m^{\prime }}(T_{1})}\frac{I_{m^{\prime }}^{\max }(T_{1})}{
I_{m^{\prime }}^{\max }(T_{2})}$ and the saturation signals cancel when $%
I_{m^{\prime }}^{\max }(T)$ is independent of $T$. When $I_{m^{\prime
}}^{\max }(T)$ does vary with $T$ it is possible to eliminate this
contribution by using an appropriate calibration method~\cite{Fotin}. Hence,
the measurement of the saturation values can be eliminated in the weak spot
regime. This is of interest when the measurement technique allows to study
the $\theta _{i}\ll 1$ range.

\begin{figure}
\includegraphics[width=10cm]{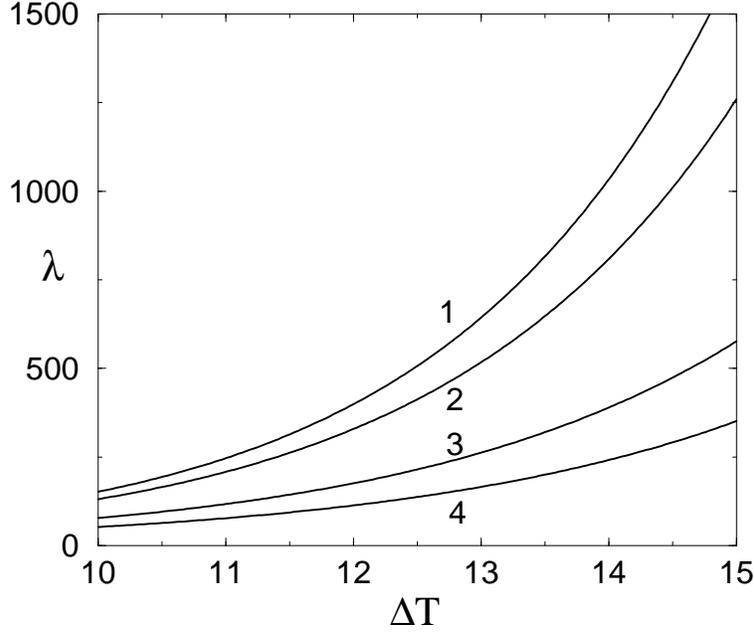}
\caption{{\bf Effect of the concentration ratio on the 
temperature dependence of 
$\Omega_{m^{\prime}}(T_{1})/\Omega_{m^{\prime}}(T_{2})$.}
$\lambda = \Omega_{m^{\prime}}(T_{1})/\Omega_{m^{\prime}}
(T_{2})$ is plotted vs. $\Delta T$ for the interval $10 
\leq \Delta T \leq 15$ for $T_{1} = 37^{\circ}C$. The concentration 
ratios are (1) $c_{m}/c_{w} = 10^{-2}$ (2) $c_{m}/c_{w}= 10^{-3}$ 
(3) $c_{m}/c_{w}= 10^{-4}$ and (4) $c_m/c_w = 0$.}
\end{figure}

\begin{figure}
\includegraphics[width=10cm]{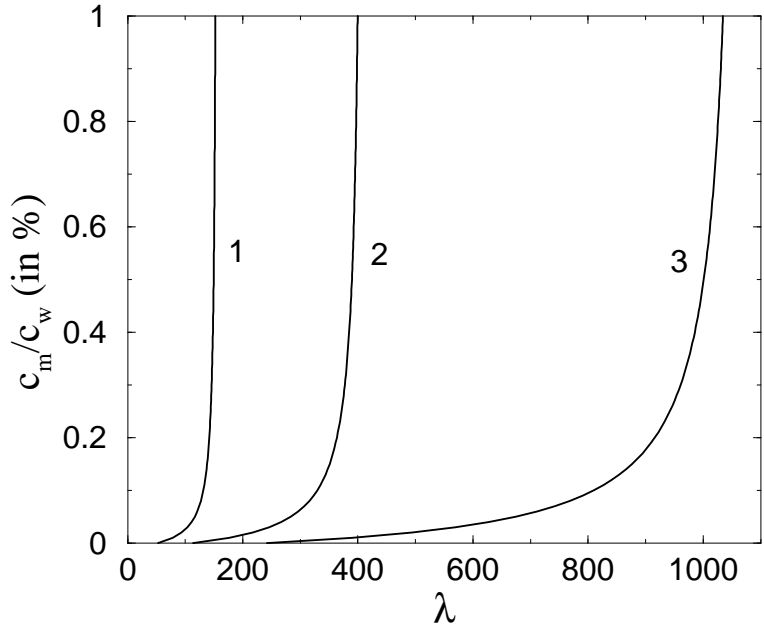}
\caption{{\bf Ratio of concentrations obtainable from 
a two temperature experiment.} 
$c_{m}/c_{w}$ vs. $\Omega_{m^{\prime}}(T_{1})/\Omega_{m^{\prime}}
(T_{2})$ for $T_1 = 37^{\circ}C$ and (1) $T_2 = 47^{\circ}C$, 
(2) $T_2 = 49^{\circ}C$ and (3) $T_2 = 51^{\circ}C$.}
\end{figure}

As we discussed, the optimization of the two spot approach is achieved by
tuning a single hybridization\ temperature. In the two temperature approach
it is attained by choosing $T_{1}$ and $\Delta T = T_2 - T_1$. The general
requirements are the same: Avoiding the saturation regimes on one hand and
the high noise to signal ratios on the other. Two additional observations
merit comments. First, increasing $\Delta T$ magnifies the range of $\lambda$
corresponding to the $c_{m}/c_{w}$ range of interest (Figure 7). Second, it
also allows to avoid the divergence regions in the $c_{m}/c_{w}$ vs. $%
\lambda $ curves (Figure 8) where ultra precise values of $\lambda$ are
required.

\section{Discussion}

Studies of somatic point mutations inevitably concern samples with a large
excess of wild type DNA. This excess is propagated by the PCR amplification
unless peptide nucleic acid (PNA) clamps~\cite{Thiede} are used. The
utilization of DNA chips to characterize the composition of such samples is
hampered by competitive surface hybridization due to the pairing of the wild
type target with the mutation probe. This mishybridization contributes
significantly to the intensity of the mutation spot signal and thus gives
rise to false positives. Our theoretical analysis suggests a systematic
approach to the minimization of such errors. At best one can extract $%
c_{m}/c_{w}$ thus obtaining additional clinical data. At least, this method
provides a rational approach for devising criteria for identification of
false positives.

DNA chips designed to detect point mutation in solid tumors were developed
by Lopez-Crapez et al~\cite{Livache}. The sample preparation in this case
involves asymmetric PCR or symmetric PCR followed by digestion with lambda
exonuclease. DNA chips of higher sensitivity for the detection of K-ras
mutation in stool were reported recently by Prix et al~\cite{Prix}. The
amplification method utilizes PNA-mediated PCR clamping~\cite{Thiede}. The
PNA binds to the wild type codons of interest thus inhibiting their
amplification. The PNA binding to the mutated DNA is weaker and does not
inhibit its amplification. The overall result is a selective amplification
of the mutated DNA. The methods we propose can be used to improve the
performance of the approach of Lopez-Crapez et al. In this case the $%
c_{m}/c_{w}$ values may provide supplementary information of diagnostic
value. In marked contrast, our methods are of little use within the approach
of Prix et al, which relies on preferential amplification of the mutated DNA
thus avoiding problems due to excess of wild type DNA.

Our analysis suggested two methods to obtain $c_{m}/c_{w}$. The advantage of
the two-spot approach is that it does not involve a change in $T$. This
approach requires however knowledge of the saturation value of the two
spots. Obtaining the saturation values can be time consuming. However, for
chips designed for multiple use this step can be performed only once
provided the saturation value does not change with the hybridization
regeneration cycles. The two-temperature approach is attractive when using a
label free detection scheme such as surface plasmon resonance. In this case
the two measurements can be carried out with no washing steps. Importantly,
Fotin et al~\cite{Fotin} demonstrated the feasibility of studying the
temperature dependence of hybridization isotherms even for targets labelled
with fluorescent tags.

Thermodynamic equilibrium is a necessary condition for the implementation of
the quantification methods proposed in this article. The equilibration times
of DNA chips and their variation with the hybridization conditions are not
yet fully understood. The reported equilibration times for hybridization on
DNA chips vary widely. In comparing the results it is important to note
differences in hybridization conditions, length of probes, type of chip and
detection techniques. Two studies are of special interest. Peterson et al
reported equilibration times of up to 14 hours~\cite{Peterson,Peterson2002}.
In marked contrast, the results of Fotin et al~\cite{Fotin} suggest
equilibration within minutes. In the present context, the work of Fotin et
al is of special interest since it reports equilibrium hybridization
isotherms at different temperatures. Importantly, the isotherms indeed
satisfy the conditions of equilibrium: Stationary state and lack of
hysteresis. Furthermore, the system exhibits a Langmuir type hybridization
isotherm and the thermodynamic parameters extracted from the temperature
dependence of the hybridization isotherms are in good agreement with the
reported bulk values. Note however that the hybridization constants
describing other types of DNA chips do not always exhibit such agreement\cite%
{Held, Zhang}.

The proposed methods require DNA chips obeying Langmuir type hybridization
isotherms. In turn, this requires spots with relatively low density of
probes. Clearly, there are advantages to chips carrying spots with high
density of probe. One is that high density enable smaller spots thus
allowing for greater number of different spots. Importantly, in DNA
microarrays the high probe density gives rise to an electrostatic
modification of the isotherms. The resulting nonlinearities are undesirable
for obtaining quantitative results.

\subsection*{Acknowledgement}
The authors benefitted from insightful discussions with T.
Livache. The work of EBZ was funded by the CNRS.

\end{document}